\documentclass[aps,prb,amssymb]{revtex4}
\usepackage{graphicx}
\usepackage{dcolumn}
\usepackage[T2A]{fontenc}
\usepackage[cp1251]{inputenc}

\begin{document}

\title{Chemical localization}
\author{V. F. Gantmakher}
\affiliation{Institute of Solid State Physics, Russian Academy of
Sciences, 142432 Chernogolovka, Moscow region, Russia}

\begin{abstract}
Analysis of experimental data shows that the metal--insulator transition
is possible in materials composed of atoms of only metallic elements. Such
a transition may occur in spite of the high concentration of valence
electrons. It requires stable atomic configurations to act as deep
potential traps absorbing dozens of valence electrons. This means in
essence that bulk metallic space transforms into an assembly of identical
quantum dots. Depending on the parameters, such a material either does
contain delocalized electrons (metal) or does not contain such electrons
(insulator). The degree of disorder is one of these parameters. Two types
of substances with such properties are discussed: liquid binary alloys
with both components being metallic, and thermodynamically stable
quasicrystals.
\end{abstract}

\maketitle


 \section{Introduction}

The band theory of metals with its concept of energy-band overlap
describes rather than explains the metallic properties of matter. The
fundamental reason for the existence of the metallic state is that in an
isolated metal atom the valence electrons occupy energy levels close to
the upper edge of the potential well, so that in the condensed state any
perturbation introduced by neighboring metal atoms leads to delocalization
of the valence electrons. From this viewpoint, the grouping chemical
elements into metals and metalloids is caused by the structure of the
atoms; metals are in the lower left corner of the Periodic Table, and the
boundary between metals and metalloids, which is a diagonal of the
Periodic Table, is blurred and extremely conventional. The transport
properties of chemical substances and substances that are a mixture of
metal and metalloid atoms depend of various factors. By selecting one of
these factors as the control parameter we can initiate a metal--insulator
transition.

The most common model describing a metal--insulator transition is the
Anderson model [1], in which disorder is the cause of the transition. The
model examines a periodic lattice of rectangular wells of different
depths. The energy levels in the wells are within an interval of values
$W$, and the level density in this interval is assumed constant. Thanks to
the tails of the wave functions $\exp(-r/\lambda)$ there is an overlap of
the wave functions of electrons localized on neighboring wells. If the
distance between the neighboring wells $r_{12}$,  is much larger than Bohr
raduis $\lambda$, i.e. if $r_{12}\gg\lambda$, the overlap integral
\begin{equation}\label{tr4}
 J\sim\int\psi_1^*\hat{H}\psi_2d^3r\sim J_0\exp(-r_{12}/\lambda)
\end{equation}
is small, with the smallness determined by the factor
$\exp(-r_{12}/\lambda)$.

Two limits are possible here. Each electron may occupy its own well
--- this is the case for very deep but different wells. On the other hand,
all electrons may be delocalized, so that any electron may find itself in
any well. For instance, if all the wells are the same or almost the same,
the electron wave functions are simply Bloch waves.

The ratio of two energies, the width of the band $W$ and the overlap
integral $J$, acts as a parameter in this problem. What Anderson stated
was that for delocalized (or extended) states to emerge, i.e. for metallic
conduction to set in, the following condition must be met:
\begin{equation}\label{tr7}
  \frac JW \geq\left(\frac JW\right)_{\rm crit}.
\end{equation}
When the ratio $J/W$ is critical, delocalized states appear in the center
of the band at $\epsilon=0$; a further increase in $J/W$ leads to a
gradual `thickening' of the layer of delocalized states.

Plugging the estimate (1) for the overlap integral into (2) and replacing
$r_{12}$ with the average distance between the centers, $n^{-1/3}$, we
arrive at the following criterion for a transition to occur:
\begin{equation}\label{Ande}
 \lambda n^{1/3}=-(\ln c_aW/J_0)^{-1},\qquad
 c_a=\left(\frac JW\right)_{\rm crit}^{-1}.
\end{equation}

We select $W$ as the measure of disorder and consider the ($W,n$) plane.
Let us assume that the Bohr radius $\lambda$ is a constant. The disorder
introduced into the system by atomic displacements has an upper limit. The
limit is reached when there is no correlation between the positions of the
atoms (we denote it by $W_{\rm max}$). In this limit the plane is reduced
to a strip (see Fig. 1). The solid curve within the strip is the
transition curve (3), with insulator states to the right of the curve and
metal states, to the left. The value $n_{\rm max}$ corresponds to the
disorder $W_{\rm max}$ on the transition curve. But what are the electron
concentrations $n$ in real metals and alloys compared to $n_{\rm max}$? If
$n>n_{\rm max}$, an Anderson transition cannot be initiated no matter how
great the disorder. Of course, $W$ can be considered a quantitative
measure of disorder only very conditionally. Hence the diagram in Fig. 1
is only an illustration. Nevertheless, the question exists and only
experiments will provide an answer.

\begin{figure}[t]
\includegraphics{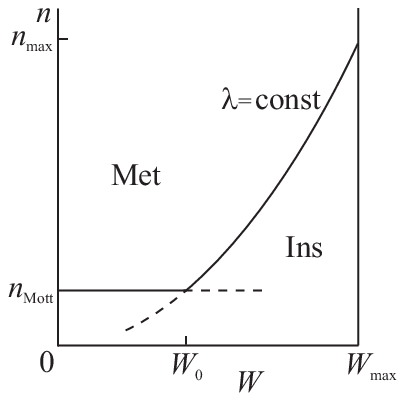}\hspace{4cm}\includegraphics{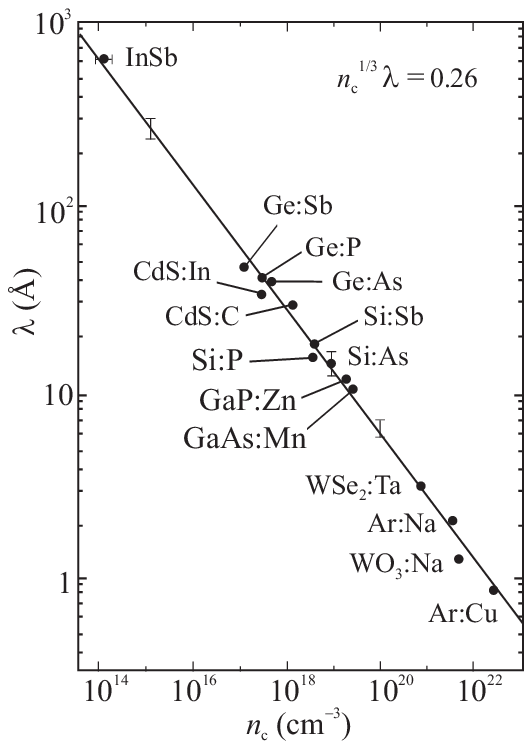}
\caption{ Diagram illustrating the possible existence of an upper limit
$n_{\rm max}$ for electron concentrations that allow for an Anderson
transition as disorder increases.} \label{mit-new2}
%
 \caption{ Correlation between the critical carrier concentration $n_c$ and
the effective Bohr radius $\lambda$ for the metal--insulator transitions
that took place in 15 different materials [3]. For all materials whose
data were used in constructing this diagram, the values of $\lambda$ and
$n_c$ were determined independently in different experiments.}
\end{figure}

Not only disorder but also the electron--electron interaction may serve as
the driving force behind a metal--insulator transition. A transition
initiated by such interaction is a Mott transition [2]. What makes it so
different from an Anderson transition is that it occurs at a fixed degree
of disorder. When the number of metal atoms is very small, even if we were
to arrange them inside an insulator to form a superlattice, so that there
is no disorder, the substance still remains an insulator. Actually, it is
impossible to vary the electron concentration $n$ and, at the same, to
keep the amplitude and the characteristic lengths of the random potential
constant; it is also impossible to vary the parameters of the random
potential and keep the concentration $n$ constant. Hence usually it is
impossible to distinguish between an Anderson transition and a Mott
transition. However, the fact that there are two possible reasons for
metal--insulator transitions should never be ignored.

The criterion of a Mott transition can easily be written in a form similar
to (3):
\begin{equation}\label{Mo}
   \lambda n^{1/3}=-(\ln c_mU/J_0)^{-1},
\end{equation}
the only difference being that instead of $W$ we have the Hubbard energy
$U$, which describes the electrostatic repulsion of two electrons
localized at a single center, and instead of $c_a$ we have a different
numerical constant $c_m$. The transition curve (4) is represented in the
diagram in Fig. 1 by a straight horizontal line, and the smaller the value
of $\lambda$ the higher the line. The two lines, (3) and (4), intersect at
the point $W_0=(c_m/c_a)U$. As long as the disorder is large ($W>W_0$),
the metal---insulator transition is controlled by it and occurs along the
curve (3). Small disorder ($W<W_0$) does not play any role since
localization occurs is caused by the electron--electron interaction at
concentration $n$ higher than those that follow from (3).

The metal--insulator transition has been realized in dozens of
experiments. But all these experiments involved systems in which the metal
atoms were diluted by nonmetal atoms, which are not inclined to provide
electrons for the general `pool' (see Fig. 2, which has been taken from
Ref. [3]). As techniques for fabricating amorphous metals (completely
disordered materials based on metal alloys) developed, it seemed that
metal--insulator transition would be discovered in them. However, even
with the enormous diversity of such alloys, their resistivity never
exceeds  $\rho^*\approx300\div400\,\mu\Omega\cdot$cm [4].

For a given concentration $n$ of carriers in the metal, $n=k_F^3/3\pi^2$,
maximum resistance occurs when the mean free path $l$ is at its minimum
$l_{\rm min}$, which is expressed in terms of the Fermi wave vector $k_F$
as follows: $l_{\rm min}\approx k_F^{-1}$. Then, to within a factor of
order unity,
\begin{equation}\label{n1}
  \rho=\frac{\hbar k_F}{ne^2l}\simeq\frac{\hbar}{e^2}k_F^{-1}(k_F l)^{-1}
  \lesssim\rho^*\equiv\frac{\hbar}{e^2}k_F^{-1}=\frac{\hbar}{e^2}n^{-1/3}.
\end{equation}

Assuming that the average atomic separation $a$ in a condensed medium is
approximately 3\,\AA, we can introduce the electron concentration in a
standard metal
\begin{equation}\label{n-st}
  n^*=a^{-3}\approx4\cdot10^{22}\,\mbox{см}^{-3},
\end{equation}
for which with a mean free path $l$ of order of the carrier separation
$n^{-1/3}$ and approximately equal to the de Broglie wavelength
$k_F^{-1}$,
\begin{equation}\label{rho*}
 \rho^*\approx\frac{\hbar}{e^2}\:\:n^{-1/3}
 \approx(200\div400)\,\mu\Omega\cdot\mbox{cm}.
\end{equation}

The maximum resistivity $\rho^*$ of a metal with such a carrier
concentration is of order of the resistivity observed in amorphous
materials. It occurs that in condensed media consisting only of metal
atoms (they are called metal alloys) a rise in disorder does not by itself
lead to localization. By introducing maximum disorder into the alloy we
only bring it closer to the brink of localization. For a transition to
occur, a fraction of the metal atoms must be replaced by metalloid atoms,
which drives the concentration $n$ of the delocalized electrons down to
below of $n_{\rm max}$.

Injection of metalloid atoms can prove to be twice as effective in the
sense that the concentration of metal atoms does not always uniquely
determine the concentration $n$ of the delocalized or potentially
delocalizable electrons. If the metal and metalloid atoms can form stable
chemical molecules, metal electrons enter the chemical bonds: from the
shallow potential well of a metal atom they go to a much deeper potential
well of the molecule and, therefore, remain localized, notwithstanding the
surroundings of the molecule. Hence the effective electron concentration
$n$, which affects the position of the material on the metal--insulator
phase diagram, decreases even more due to the emergence of chemical bonds.

Bearing in mind the tying-up of a fraction of the potentially free valence
electrons into chemical bonds, we can formulate the following question:
{\it Is there a way to build deep potential wells using only metal atoms,
which would transform a material in which there are no metalloid atoms
into an insulator with an electron concentration (6) of a standard metal?}
The experimental data discussed in the present paper shows that this is
possible.

\section{Intermetallic compounds in two-component melts}

For a long time it has been known that the resistivity of a liquid melt of
two very good metals may change severalfold, even by a factor of 10,
depending on the relative concentration of the two components, and reach
its maximum at a certain rational ratio of the atomic concentrations, such
as 1:1 or 1:3 or 1:4 [5, 6]. To gather such data one must know how to
measure the resistivity at a fixed temperature as a function of the alloy
component concentration. A description of the respective experimental
facilities can be found in Refs. [7, 8].
\begin{figure}[t]
\includegraphics{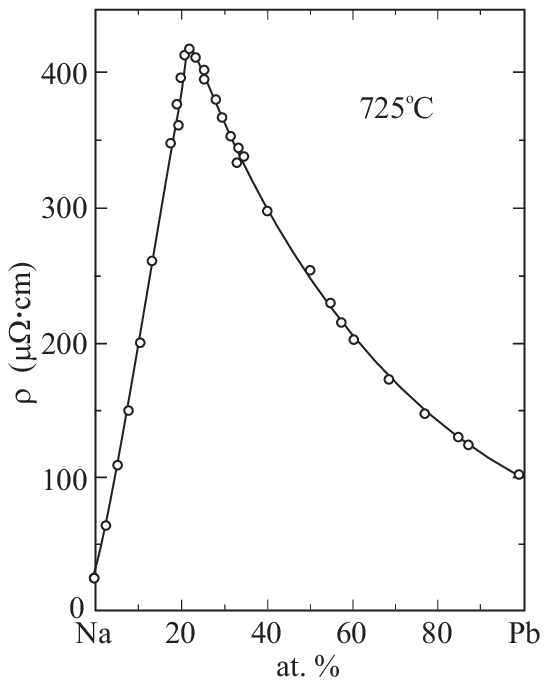}\hspace{4cm}\includegraphics{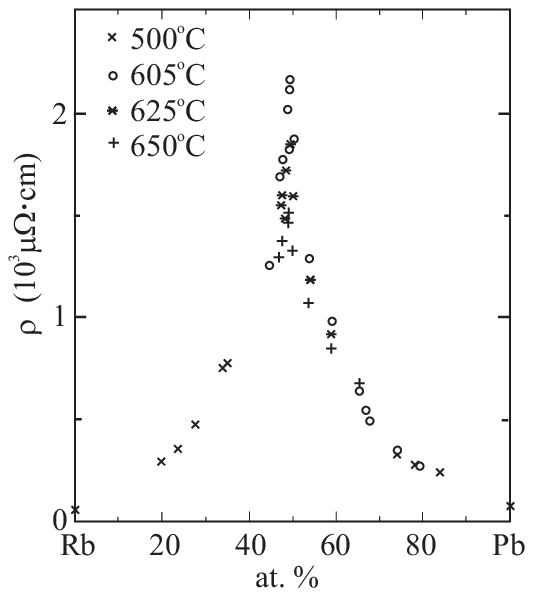}
\caption{ Resistivity of melts of the Na--Pb system at 725$^\circ$C. The
peak value is reached at a lead concentration $C_{\rm Pb}$=20\%, where
stable PbNa$_4$ configurations emerge [8].}
 \caption{ Resistivity of melts of the Rb--Pb system at different
temperatures [9]. The peak value is reached at a lead concentration
$C_{\rm Pb}$=50\%; the Pb$_4$Rb$_4$ configurations are stable.}
\end{figure}

\begin{figure}[b]
\includegraphics{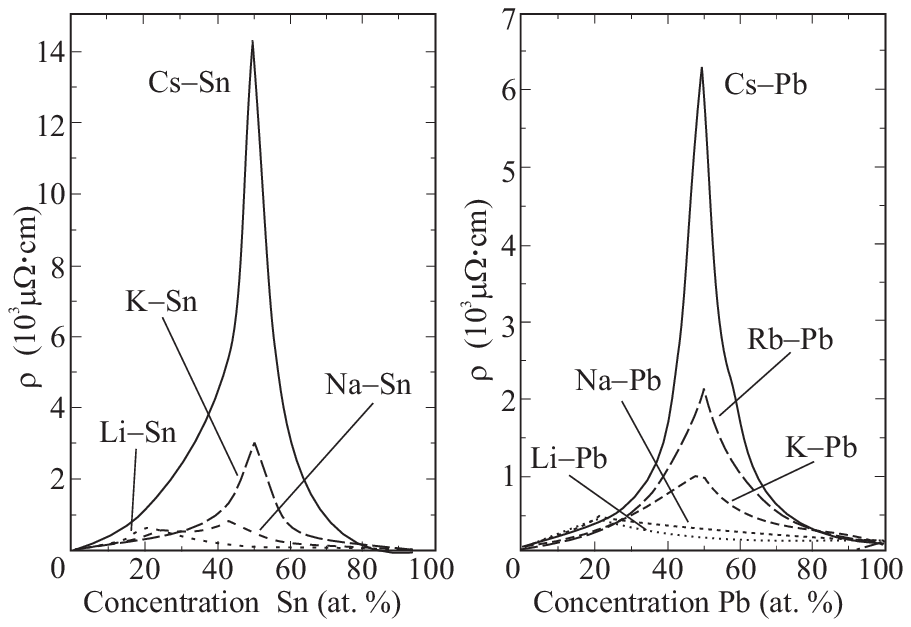}
\caption{Survey diagrams of the resistivity vs. concentration dependence
for melts of Sn--B and Pb--B systems (B is an alkali metal). The
resistivity of alloys with Li and the Pb--Na alloy has a maximum at
$C_{\rm A}$=20\% (A=Pb,\,Sn); the resistivity of the Sn--Na alloy has two
maxima at $C_{\rm Sn}\approx$\,25\% and 45\%; the rest have a resistivity
maximum at $C_{\rm A}$=50\% [10].}
\end{figure}

Figure 3 shows the results of measurements of the resistivity of Na--Pb
melts at 725$^\circ$C done by Calaway and Saboungi [8]. Clearly, the
concentration ratio Na:Pb$\sim$4:1 is preferred. An addition of 20\% of
lead increases the resistivity compared to that of pure Na by a factor of
about 20. Here the peak value of resistivity is of order of the maximum
possible value $\rho^*$ of a standard metal. The Li--Pb system behaves in
a similar manner.

Replacing Li and Na with a heavier alkali metal, K, Rn, or Cs, changes the
resistivity vs. concentration diagram significantly. For example, Fig. 4
shows the diagram for the Rb--Pb system taken from Ref. [9]. The peak has
shifted to another rational ratio of the component concentrations
Rb:Pb$\sim$1:1, while the peak value of the resistivity increased
severalfold. Now this value exceeds the maximum resistivity (7) of a
standard metal (6) by a factor of 10. The survey diagrams in Fig. 5 show
that melts of alkali metals with another tetravalent metal, tin, behave in
the same manner [10]. The highest resistivity values are realized in
Cs-based melts.

The high values of resistivity mean that near the respective concentration
ratios the melt ceases to be a standard metal in the sense that a fraction
of carriers in it are bound in some manner and the remaining effective
concentration $n_\mathrm{eff}\ll4\cdot10^{22} \mbox{cm}^{-3}$ (cf. (6)).
Indeed, for the Rb--Pb system, the value
$\rho\approx2200\,\mu\Omega\cdot$cm is 10 times greater than the maximum
value for a standard metal, $\rho^*\approx(200\div400)\,\mu\Omega\cdot$cm.
According to equations (5) and (7), this implies that the number of free
carriers in the melt is no greater than $10^{-2}\div10^{-3}$ of the
ordinary number of carriers in a standard metal.

From the rational component-concentration ratios it follows that the
increase in resistivity is due to formation of compounds within which
most electrons prove to be locked. The position of the peak in
resistivity in Li- and Na-based melts unquestionably points to the
existence of Na$_4$Pb and Li$_4$Pb compounds in the melts. The five atoms
comprising such a compound have eight electrons in their valence shells.
Apparently, they form a single stable outer shell of the Pb$^{4-}$ ion,
while four alkali ions held together by Coulomb forces surround that ion.
The four ions form a barrier thanks to which the eight electrons in the
outer shell of Pb are kept within this electrically neutral atomic
configuration and do not participate in conduction (Fig. 6a).
\begin{figure}[h]
\includegraphics{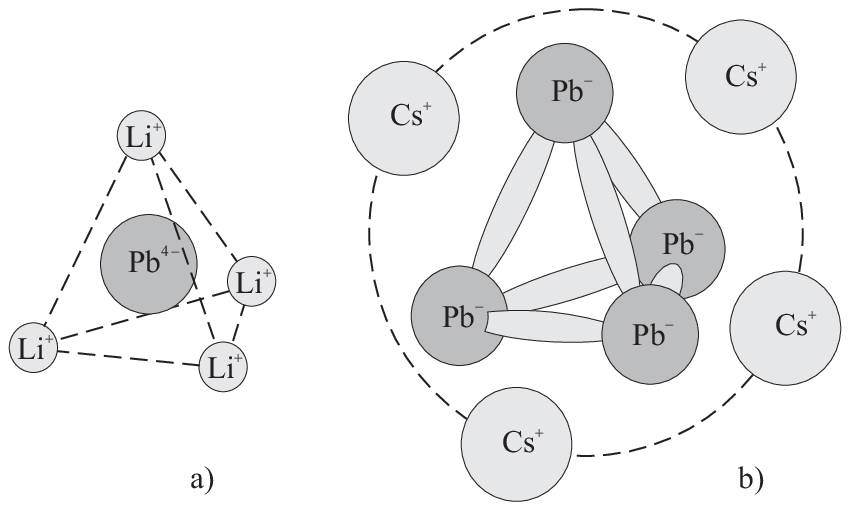}\hspace{3cm}\includegraphics{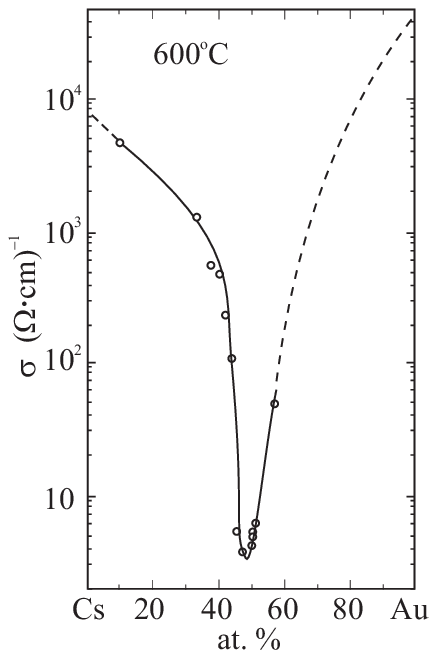}
\caption{ (a) Ionic configurations from Pb and Sn atoms and a light
alkali metal Li or Na; (b) the same with a heavy alkali metal K, Rb, or
Cs. }
 \caption{ Conductivity of melts of the Cs--Au system at
600$^\circ$C [5, 12]}
\end{figure}

An increase in the size of the alkali atoms leads to a qualitative change
of the forming compounds, with the ability of these compounds to act as
electron traps gaining in strength. Such structures are well known and are
called Zintl's structural units (named after the German chemist who in the
1930s discovered the rule of formation of ionic configurations [11]). If
an electron goes from an alkali atom to the lead atom, the Pb$^{1-}$ ion
will have five electrons in the outer shell, the same as in the P or As
atoms. As is known, these two elements form in the gaseous phase
tetrahedral molecules P$_4$ or As$_4$. Here there are eight electrons near
each atom: five electrons belonging to the atom proper and one electron
from the covalent bonds with each of the three neighbors in the
tetrahedron. Pb$^{1-}$ ions also form such tetrahedrons, and the total
electric charge $-4e$ of such a tetrahedron is balanced by the electric
charge of the four alkali-metal ions surrounding it. Sn$^{1-}$ ions form
similar tetrahedrons (Sn$_4)^{4-}$ surrounded by four alkali ions. It is
the structural unit
\begin{equation}\label{AB}
  {\rm A_4B_4,\qquad A=Pb,\:Sn,\qquad B=K,\:Pb,\:Cs}
\end{equation}
that is the configuration within which 20 valent electrons are locked
(Fig. 6b).

Binary melts consisting of alkali metals and some other metals behave in a
similar manner. The absolute champion when it comes to forming effective
electron traps is the isoatomic melt of two ideal metals, the alkali metal
Cs and the noble metal Au. As Fig. 7 shows, the formation of compounds in
the melt reduces the conductivity by a factor of 10000 [12]. Here the
conductivity is comparable to that of salt melts
(3\,$\Omega^{-1}\cdot$cm$^{-1}$ for CsAu and
1\,$\Omega^{-1}\cdot$cm$^{-1}$ for the CsCl salt melt).

\section{The model with structural disorder}

Quantum chemistry and chemical thermodynamics have the tools that are
needed to answer the questions of where, when, and how many Zintl
configurations can form in a metallic melt and what are the binding
energies of these configurations. Since melts, by definition, exist at
high temperatures, the curves in Figs. 4 and 7 should not be considered as
demonstration of metal--insulator transitions. To bring these materials
into the realm of objects described by the theory of metal--insulator
transitions, they should be quenched into glass. Then, for example, the
low-temperature dependence of the transport characteristics can, probably,
be used to determine the quantitative parameters of the electron traps. So
far nothing is known of any attempts to quench such melts with a view to
investigating their low-temperature properties.

At the same it is true that what occurs when component concentrations are
stoichiometric is undoubtedly localization. Therefore, when discussing the
results of experiments in this area of research, it seems appropriate to
use the concepts and models developed for a description of
metal--insulator transitions.

\begin{figure}[h]
\includegraphics{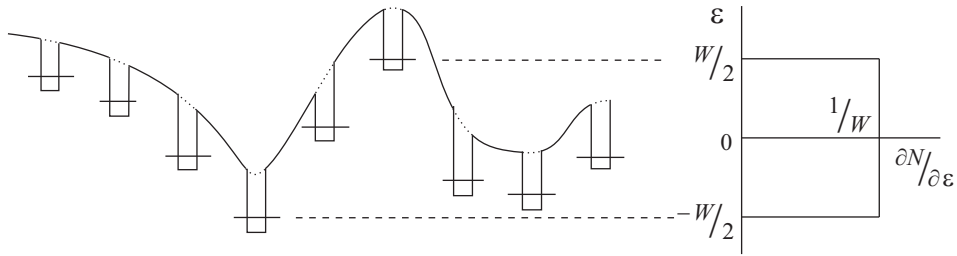}
\caption{ The Anderson model: periodically arranged wells against random
smoth potential (to the right is the model density of states).}
\end{figure}

Earlier we mentioned the most common model of regularly arranged potential
wells of different depths, or the Anderson model (Fig. 8). Each well is
formed because of the potential of one or several atoms. For the levels in
the wells to form a continuous band, the wells must be arranged against a
background of smooth random fields, say, the electric fields of charged
acceptors or donors with partial compensation of impurities (see Chap. 3
in Ref. [13]). Such broadening of a level into a band may be called
classical.

The alternative of the Anderson model for the description of a transition
in a system of noninteracting electrons that is driven by a change of
disorder is the model with structural disorder. Here the random potential
is built from identical but randomly distributed wells each of which
contains a level $E_0$
\begin{equation}\label{trz1}
  V({\bf r})=\sum_{{\bf R}_i}v({\bf r}-{\bf R}_i),
\end{equation}
with the disorder determined by the randomness of the set of vectors ${\bf
R}_i$. The model contains no classical random fields. Despite the fact
that all the wells are identical, the level $E_0$ in this model also
broadens into a band. The broadening is caused by the quantum interaction
of the wells due to the overlap of their wave functions [14].

\begin{figure}[b]
\includegraphics{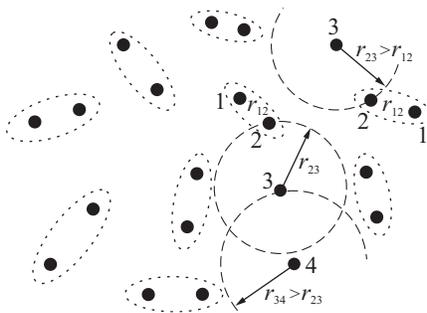}
\caption{Random arrangement of the potential wells. Pairs of nearest
neighbors are designated by dashed ellipses. Centers labelled 3 adjacent
to pairs whose wells are labelled 1 and 2 form triplets with these pairs.
One of the triplets enters into a quadruplet.}
\end{figure}

We now break down all the pairs into nearest neighbors. If the distance
between the wells in such a pair is $r_{12}$, then, since the wells are
resonant, i.e. the unperturbed values of the level energy are the same,
the overlap of the tails of the wave functions splits the levels into two
levels with energies
\begin{equation}\label{split1}
  E=E_0\pm\varepsilon_{1,2},\qquad
  \varepsilon_{1,2}=J_0\frac{\exp(-r_{12}/\lambda)}{r_{12}}\:.
\end{equation}
They have collectivized wave functions
\begin{equation}\label{split2}
\psi_{1,2}=\frac{1}{\sqrt{2}}(\varphi_1\pm\varphi_2),
\end{equation}
expressed in terms of the unperturbed wave functions $\varphi_1$ and
$\varphi_2$.

The constants $J_0$ and $\lambda$ contain the specific characteristics of
the wells, the dielectric constant of the material, the effective electron
mass, etc. Due to the splitting (10), both levels $\varepsilon_{1,2}$ are
no longer resonant with respect to other neighboring levels, and their
interaction with such levels leads to essentially smaller energy shifts
\begin{equation}\label{split3}
  \Delta\epsilon\sim\exp(-2r_{s,t}/\lambda)\qquad
  (s\quad\mbox{или}\quad t\:\neq1,2).
\end{equation}

Figure 9, where the resonant pairs are designated by dashed ellipses,
shows that not all centers belong to resonant pairs. For instance, well 2,
being the nearest neighbor of well 3, may have well 1 as its nearest
neighbor, so that $r_{12}<r_{23}$. Within this triplet, the resonant
shifts $\varepsilon_1$ and $\varepsilon_2$ are the biggest, while the
shift $\varepsilon_3$ is nonresonant and much smaller, since
$\varepsilon_3\propto\exp(-2r_{23}/\lambda)$. Two such configurations have
been depicted in Fig. 9, a triplet and a quadruplet, the latter with
$r_{12}<r_{23}<r_{34}$. But in triplets and more complicated
configurations consisting of four and more wells there is always at least
one resonant pair with the smallest well separation and the biggest level
shift [14]. Since the characteristic width $\Delta$ of the resulting
density of states is determined by the resonant well pairs and the average
well separation is $n^{-1/3}$, from (10) we obtain
\begin{equation}\label{trz3}
  \Delta\approx J_0n^{1/3}\exp(-n^{-1/3}/\lambda)\:.
\end{equation}

The tail of the density of states in the region $|\varepsilon|\gg\Delta$
emerges due to the pairs of anomalously close wells with $r_{12}\ll
n^{-1/3}$, while states with small $|\varepsilon|\ll\Delta$  emerge due to
nonresonant and single wells ([14]; see also Chap. 2 in Ref. [13]).
Formula (13) clearly shows that the ratio of the decay length $\lambda$ to
the average distance $n^{-1/3}$ between wells is the main parameter in the
structural disorder model, just as it is in the Anderson model. (Note: In
the metallic limit $\lambda n^{1/3}{\gg}1$, where all the electrons are
delocalized (i.e. collectivized) and the potential (9) is screened and is
only a scattering source, this potential is used in the diffraction theory
of electron transport in liquid metals (see Ref. [15])).

The metal--insulator transition in the region where
\begin{equation}\label{param}
  n^{1/3}\lambda\sim1
\end{equation}
has been studied theoretically much less thoroughly in the model with
structural disorder than in the Anderson model and, apparently, only
numerically (e g see Ref. [16]). Still, in this model such a transition
undoubtedly exists. Since initially all the ionic traps, or Zintl
configurations, are identical, the model with structural disorder seems to
be more appropriate for describing them. Each configuration A$_4$B$_4$
from (8) is an almost spherical well for the $4{\cdot}4+4=20$ valence
electrons located inside it on a sequence of energy levels [17]. The decay
length $\lambda$ in (14) actually refers to the uppermost occupied level.
The electrons on deeper levels do not leave the well. This reduces by a
factor of 10 the concentration of potentially delocalizable electrons and
facilitates the metal--insulator transition.
\begin{figure}[h]
\includegraphics{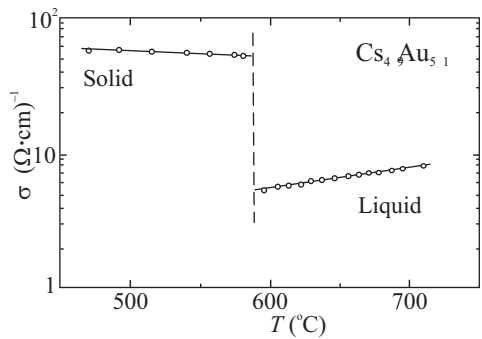}
 \caption{Temperature dependence of the conductivity of the CsAu alloy
with 51\% of Au in the liquid and solid states [5, 7].}
\end{figure}

Thus, the number of electrons and the level structure in an ionic trap
determine by how much the concentration $n$ in the parameter (14) of the
model with structural disorder is reduced, while the shape of the well
$v({\bf r}-{\bf R}_i)$ determines $\lambda$. Another control parameter in
this model is the magnitude of correlations on the set of vectors ${\bf
R}_i$ : by strengthening the correlations one can change this set from
random to regular. The result is shown in Fig. 10, which depicts the
change in conductivity of an almost stoichiometric alloy CsAu under
crystallization [5, 7]. Here the majority of the wells become resonant and
have identical and identically located neighbors. The randomness in the
location of the wells is partially retained only to the extent to which
the alloy is nonstoichiometric and due to the presence of crystal defects
and intercrystalline boundaries. The result of the increase in the number
of resonant wells is partial delocalization and a tenfold increase in the
conductivity of the crystal compared to that of the melt. However, as Fig.
10 clearly shows, the conductivity of crystalline CsAu is still about 50
times lower than the maximum conductivity $1/\rho^*$ of a standard metal
(equation (7)). It remains unclear to what extent and how the conductivity
depends on deviations from stoichiometry, the number of defects,
temperature, and other factors.

\section{Quasicrystals}

Introducing translational symmetry is not the only way to establish
long-range correlations on the set of vectors ${\bf R}_i$. Another way to
achieve the same result is to see to it that quasicrystalline long-range
order sets in [18].
\begin{figure}[t]
\includegraphics{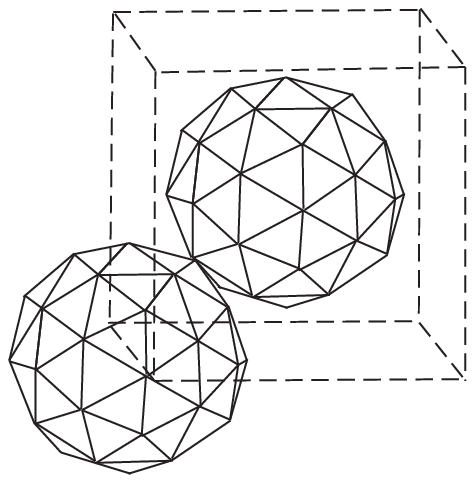}
\caption{ A crystal packing of Mackey icosahedrons, closely resembling a
body-centered cubic packing, in the crystal alloy $\alpha$(AlMnSi), a
quasicrystal approximant [19]. Each icosahedron consists of more than 50
atoms.}
\end{figure}

Translational symmetry, always present in crystals, allows for the
existence of axes of 2-, 3-, 4-, and 6-fold symmetry only. At the same
time, it is easy to imagine that of all possible local configurations of a
small number of atoms, A$_n$B$_m$C$_p$, of the chemical elements A, B, and
C the configuration with the lowest energy has a different symmetry axis,
say the axis of 5-fold symmetry. Formation of a crystal from a material
with the composition A$_n$B$_m$C$_p$ then becomes a problem. Sometimes
optimal local symmetry is sacrificed, so that a crystal with a different
configuration of the nearest neighbors of each atom is formed, with the
loss in local configurations energy balanced by gain caused by
translational symmetry. There is, however, another possibility. Let us
arrange optimal configurations of $n+m+p$ atoms at the sites of a crystal
lattice, say, a body-centered cube, as in Fig. 11. Then the loss in energy
emerges caused by mismatch and distortions in the places where these
configurations meet, where the short-range order is sure to be nonoptimal.
Nevertheless, some substances have such crystal structures. They are
called crystal approximants or crystal prototypes of quasicrystals.
\begin{figure}[b]
\includegraphics{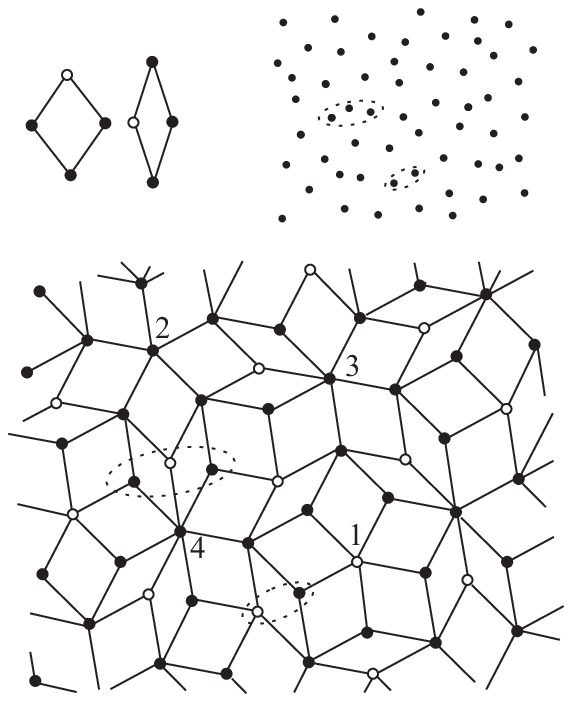}\hspace{5cm}\includegraphics{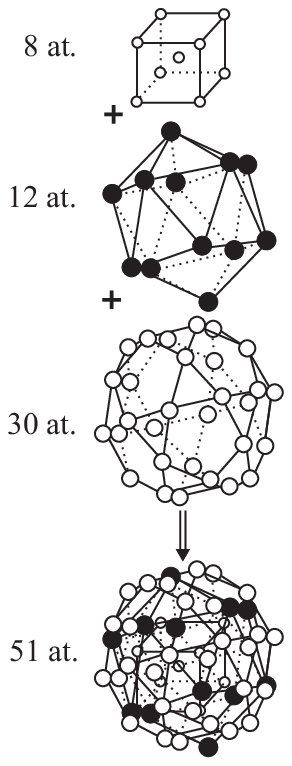}
\caption{ Penrose tiling. Below: tiling a plane without gaps or overlaps
by two types of rhombic tiles depicted in the upper left corner of the
figure, rhombuses with equal sides $a$ and acute angles 2$\pi$/5 and
$\pi$/5, respectively (the vertices marked by open circles adjoin each
other in the tiling). Upper right corner: the set of rhombus vertices of
the tiling depicted in the lower part of the figure to the 1:2 scale.
Although each of the sites 1, 2, 3, and 4 has its nearest neighbors only
at a distance $a$, the quality of these neighbors differ considerably
(see the main text). The dotted closed curves mark a resonant pair of
closely located sites and a compact triplet of sites.}
 \caption{A sequence of shells consisting of atoms in a Mackey
pseudoicosahedron, which is the base element of the structure of the
Al--Pd--Mn quasicrystal [20]; the total number of atoms is 51.}
\end{figure}

It occurs, however, that we can do entirely without translational symmetry
by densely packing the space with optimal configurations. That this is
possible, at least theoretically, is demonstrated by the Penrose tiling in
the lower part of Fig. 12; the plane is covered perfectly (i.e. without
gaps and overlaps) by rhombic tiles of two types, with the acute angles
equaling 2$\pi$/5 and $\pi$/5. Single tiles are depicted in the upper left
corner of the same figure. Since the rhombuses adjoin each other at
preassigned vertices, the correctly specified functions $F({\bf r})$ on
the rhombuses remain continuous at the junctions and form a continuous
aperiodic function whose separate segments are repeated in the plane an
infinite number of times. In the upper right corner of Fig. 12 the set of
the vertices of the rhombuses are depicted to the 1:2 scale. Since there
is no translational symmetry, it is rather difficult to notice any
correlations in this arrangement. However, there is long-range order in
this system: the rhombic tiles are arranged on the plane in a
well-distinguishable pattern (although the pattern is not unique).

Quasicrystals are built according to the same principles. Optimal
configurations with high-order symmetry axes are separated by matching
configurations-spacers that minimize energy losses at the junctions. The
resulting arrangement has no translational symmetry but does not have
long-range order. Many families of such materials are known today. Most of
them are metal alloys in the sense that they consist only of metal atoms:
Al--Mn, Ga--Mg--Zn, Al--Cu--Fe, Al--Pd--Re, etc. Here the local base
configurations may be extremely complex. For instance, in quasicrystals
with the Al--Pd--Mn composition one of the local base configurations
consists of three sells inserted into each other [20]; altogether there
are 51 atoms in this configuration (Fig. 13).

Basically, quasicrystals are identified and studied by the X-ray
diffraction method. The Fourier transform of any function of coordinates
in a perfect crystal, e.g. the density $\varrho(\bf r)$, is a sum of an
infinite number of narrow peaks (ideally, $\delta$-functions):
\begin{equation}\label{ch2}
  \varrho(\bf r)=\sum_q\varrho_q\exp({\it i}\mathbf{qr}).
\end{equation}
The set of vectors $\bf q$ form a lattice in the $q$-space with the same
symmetry as the initial lattice of atoms. To each site of this reciprocal
lattice there corresponds a Bragg reflection in the Laue diffraction
pattern. The more perfect the crystal the shaper the reflections.

Bragg reflections are not an exceptional property of crystals. As the
Fourier transform we can initially select the series (15) in which the set
of vectors $\bf q$ does not possess translational symmetry. By an inverse
Fourier transformation we arrive at a function $\varrho(\bf r)$ that has
no translational symmetry either. Such series are the Fourier transforms
of quasicrystals. Here the width of the Bragg reflections is still
determined by the imperfectness of the structure, namely, by deviation of
the local configurations from the ideal configuration, failure of
long-range order because of impurities and vacancies, etc. The sharper the
Bragg reflections the closer the quasicrystal is to a perfect one.

The following correlation is typical for metallic single crystals: the
higher the quality of the Laue diffraction pattern of a certain substance,
the lower the residual resistivity $\rho$ of the crystals. The correlation
reflects the wave nature of electrons: the better the conditions for the
propagation of an X-ray wave, the smaller the scattering of the Bloch
wave. In quasicrystals it is just the opposite: annealing, while
increasing the quality of the Laue diffraction pattern, also increases the
resistivity. Here the very values of resistivity are extremely high [18].
For instance, in quasicrystals of the Al--Cu--Ru composition at 4 K the
resistivity values are as high as 30\,$m\Omega\cdot$cm, which is
approximately a 100 times higher than the value of $\rho^*$ estimated by
equation (5) from the concentration $n$ of the metallic valence electrons.

The properties of an insulator manifest themselves most vividly for the
Al--Pd--Re system, where the resistivity values at 4 K are stably of order
$200\div300\,m\Omega\cdot$cm [18]. Ingots of this alloy can be made by arc
melting a mixture of extremely pure Al, Pd, and Re in an atmosphere of
pure argon. After being annealed in vacuum for 24 hours at 980$^\circ$C
the alloy becomes an icosahedral quasicrystal. However, even after this it
remains sensitive to low-temperature annealing at 600$^\circ$C. Such
annealing in the course one to two hours may double or even triple the
resistivity at 4 K, with the quality of the Laue diffraction pattern
remaining the same or even growing.

\begin{figure}[b]
\includegraphics{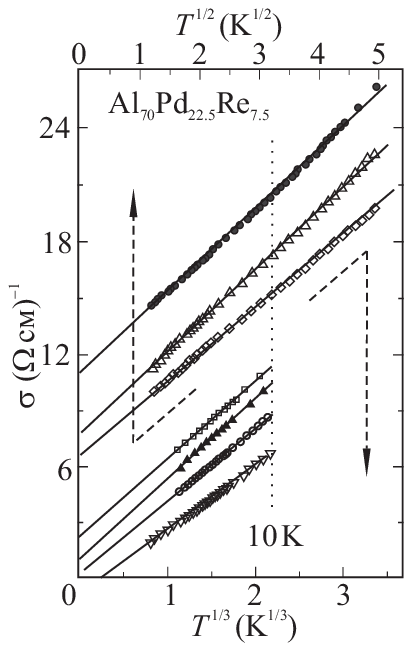}\hspace{4cm}\includegraphics{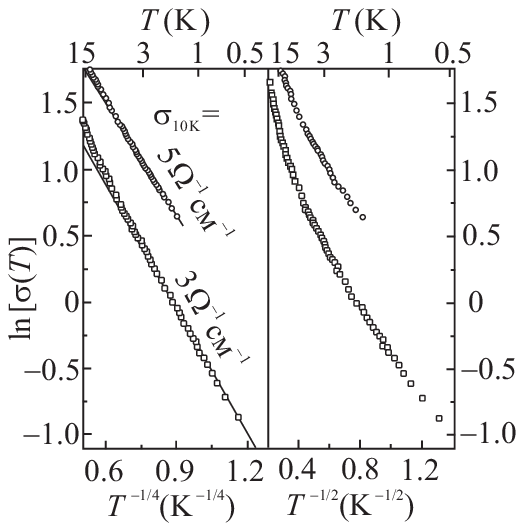}
\caption{Temperature dependence of the conductivity of the
Al$_{70}$Pd$_{22.5}$Re$_{7.5}$ quasicrystal. In the immediate vicinity of
the metal--insulator transition, the dependence, when represented by a
function of $T^{1/3}$, is a straight line (the four lower states). In the
bulk of the metallic region, the dependence becomes a straight line when
represented by a function of $T^{1/2}$ (the three upper states). The
states can be labelled by the magnitude of the conductivity $\sigma_{10}$
at 10 K. (The data has been taken from Ref. [21])}
 \caption{Mott's law for the conductivity of Al$_{70}$Pd$_{22.5}$Re$_{7.5}$
quasicrystals in the insulator region (the conductivities at 10 K are 5
and 3 $(\Omega\mbox{см})^{-1}$). The dependence becomes a straight line
only when $\ln\sigma$ is plotted as a function of $T^{-1/4}$ [27].}
\end{figure}

The temperature dependence of the resistivity of
Al$_{70}$Pd$_{22.5}$Re$_{7.5}$ quasicrystals can be described in exact
agreement with the existing theoretical scheme used to describe conduction
in the vicinity of metal--insulator transitions in ordinary media. Figure
14 depicts the conductivity as a function of $T^{1/3}$ (the four lower
curves) or $T^{1/2}$ (the three upper curves) [21, 22]. To distinguish
between the various samples and between the various states of a single
sample obtained in the low-temperature annealing process, we select the
value $\sigma_{10}$ of conductivity at 10 K as a parameter (the scales
along the horizontal axes in Fig. 14 have been selected in such a way that
at this temperature the two scales coincide, as they also do at $T=0$).

Figure 14 clearly shows that for all measured functions $\sigma(T)$ a
linear extrapolation on the selected scales makes it possible to determine
$\sigma(0)$ (cf. Ref. [23]). For the tree upper states with
$\sigma(0)\gtrsim6(\Omega\mbox{cm})^{-1}$ we can assume that
\begin{equation}\label{ch2a}
  \Delta\sigma\simeq\sigma_{10}-\sigma(0)\lesssim\sigma(0).
\end{equation}
This makes it possible to assume that the temperature-dependent part of
the conductivity is a quantum correction [24], and this is why the
function $\sigma(T)$ looks as a straight line in the $(T^{1/2},\sigma)$
plane. For the four lower states with
$\sigma_{10}\lesssim12\div14\,(\Omega\mbox{cm})^{-1}$ we have the opposite
of (16). This means that these states are in the critical vicinity of the
metal--insulator transition. Hence, when built in the $(T^{1/3},\sigma)$
plane, the function $\sigma(T)$ is represented by a straight line [24,
25]:
\begin{equation}\label{ch3}
 \Delta\sigma\equiv\sigma(T)-\sigma(0)\propto T^{1/3}.
\end{equation}
The function $\sigma(T)$ in the critical region in the vicinity of the
metal-insulator transition should evolve in this manner (e.g. see Ref.
[26]).

Correspondence to ordinary behavior is retained in states with higher
resistivity values. The extrapolation we have described implies that the
metal--insulator transition occurs at the state with
$\sigma(10\,K)\equiv\sigma_{10}\simeq9\,(\Omega\mbox{cm})^{-1}$. For
states with smaller values of $\sigma_{10}$ low-temperature transport is
realized through the hopping conduction mechanism. That this is actually
the case is illustrated by Fig. 15 taken from Ref. [27]. The diagram
shows that the conductivity of high-resistance
Al$_{70}$Pd$_{22.5}$Re$_{7.5}$ quasicrystals obeys Mott's law
\begin{equation}\label{ch4}
  \ln\sigma\sim T^{-1/4}.
\end{equation}
Such temperature dependence implies that near the Fermi level the density
of states of the electronic spectrum has a constant, finite value.

Thus, everything that happens with Al$_{70}$Pd$_{22.5}$Re$_{7.5}$
quasicrystals under low-temperature annealing, which improves conditions
for the propagation of electromagnetic wave packets, fully corresponds to
the pattern of metal--insulator transitions as the parameter
$n^{1/3}\lambda$ decreases.

Although the tendency of the resistivity to increase as the Bragg
reflections get narrower is a characteristic feature of many families of
quasicrystals, so far the metal--insulator transition has been observed
only in the Al--Pd--Re. The maximum resistivity values at 4 K for the
Al--Cu--Fe, Al--Cu--Ru, and Al--Cu--Mn systems are smaller than the value
for Al--Pd--Re by factor of 10 to 100. Note that even these values are
higher than $\rho^*$ calculated by (7) by a factor of 10 to 100.

Let us try to understand how the insulator Al$_{70}$Pd$_{22.5}$Re$_{7.5}$
is organized. The Al$_{70}$Pd$_{22}$Mn$_8$ structure is the most
thoroughly studied one and differs from Al$_{70}$Pd$_{22.5}$Re$_{7.5}$ in
only one aspect, i.e. the isovalence Re is replaced by Mn. The
quantitative characteristics of these quasicrystals can be assumed to be
the same.

Thus, the structure of the Al$_{70}$Pd$_{22}$Mn$_8$ and
Al$_{70}$Pd$_{22.5}$Re$_{7.5}$ quasicrystals is based on high-symmetry,
close to spherical configurations consisting of 51 atoms (see Fig. 13).
According to diffraction data, the number density of the atoms in these
substances is close to $6{\cdot}10^{22} \mbox{cm}^{-3}$. Since the atoms
of the transition elements `grab' some of the three valence electrons of
aluminum, the number of the remaining `potentially metallic' electrons is
somewhat smaller than two per atom, i.e. about $10^{23}\mbox{cm}^{-3}$.
For a substance with such a huge electron concentration to be an
insulator, the electrons must reside in deep potential wells, or traps. In
intermetallic binary melts the traps are configurations of type (8), while
in quasicrystals they are the high-symmetry configuration of Fig. 13,
which have levels for about 90 of the former valence electrons [20]. Under
favorable condition only one to two electron from the upper levels may
leave a trap. Hence initially the electron configuration is reduced by a
factor of 100, after which the more or less standard models describing the
metal--insulator transition can be employed.

The arrangement of the levels in all atomic configurations that are traps
is, in the zeroth approximation, the same. If the configurations were
arranged periodically, the levels would become (in accordance with band
theory) bands and the electrons from the upper levels could become
delocalized. However, in a quasicrystal there are many ways in which the
neighboring configurations can be arranged in relation to a given
configuration. According to the model with structural disorder, each
variant of the surroundings corresponds to a specific shift of the levels
in the given configuration. Let us explain this using Penrose tiling as an
example, for which we turn to Fig. 12. The distance between a given site
and a neighboring site can be equal to the length $a$ of the rhombus side
or to the length of the smaller diagonal of the narrow rhombus,
$a_1=0.62a$, or to the length of the smaller diagonal of the wide rhombus,
$a_2=1.18a$. However, the number of variants of the surroundings, which
determine the shift of the level of a specific site, is very large. For
instance, sites with close neighbors at distance $a_1$ may form resonant
pairs or triplets. Sites 1 and 2 each have five neighbors at a distance
$a$, but all five neighbors of site 1 enter into resonant pairs or compact
triplets with pairwise distances $a_1<a$, while site 2 has no such
neighbors; site 3 has six neighbors at a distance $a$, but three of these
neighbors form a compact triplet; site 4 has seven neighbors at a distance
$a$, but six of these neighbors form two compact triplets; etc. As a
result, a single level, which initially was the same for all
configurations (sites), becomes a band. Whether or not the states in this
band are localized depends on the parameter (14), where the $\lambda$ is
the decay length of the wave function outside the configuration well.

The very fact that the conductivity of Al$_{70}$Pd$_{22.5}$Re$_{7.5}$ is
so small is, apparently, caused by the specific combination of the
parameters of the configuration well, which makes the decay length
$\lambda$ smaller than in other quasicrystals. In the event of
low-temperature annealing of Al$_{70}$Pd$_{22.5}$Re$_{7.5}$, the
configuration wells in the quasicrystal undergo `internally repair'
accompanied by a decrease in the leakage of the wave function from the
well, i.e. a decrease in the effective decay length $\lambda$.

\section{Conclusion}

Processes that form the electronic spectrum in two-component melts with an
alkali metal as one of the components and in quasicrystals have proved to
be very similar. In such systems the effective carrier concentration
decreases and the screening weakens while material structuring, which
makes an ordinary metal--insulator transition possible.

The overall scheme is as follows. Suppose that each configuration contains
$N$ valence electrons. The potential produced by the ion cores of the
atoms in a configuration is so strong, i.e. the potential well is so deep,
that the electronic spectrum of these $N$ electrons becomes radically
transformed so that the electrons occupy `positions' on a ladder of
levels. Only one or two electrons on the upper levels have a chance of
leaving the well. As a result, the concentration of `potentially
delocalizable' electrons becomes of order $n/N$, where $n$ is the
concentration of the `initially metallic' valence electrons. In
two-component melts, $N$ is of order 10, while in quasicrystals it is of
order 100. Thus, the metal--insulator transition occurs in a system with a
reduced carrier concentration.

The same line of reasoning can be formulated differently if we imagine
each configuration as being a quantum dot in 3D space. The concentration
of such dots is of order $n/N$, with each dot containing $N$ electrons.
When one electron leaves a quantum dot, the dot's charge increases by $e$,
and this requires energy of order
\begin{equation}\label{Coul}
  \varepsilon_e\approx e^2/r,
\end{equation}
where $r$ is the radius of the quantum dot. This quantity is similar to
the Hubbard energy in the theory of Mott's transitions [2]. At the same
time, $e^2/r$ is the Coulomb energy of an isolated metal sphere of radius
$r$ carrying charge $e$ or the energy of the capacitor that appears in the
theory of the Coulomb blockade in nanostructures. On the metal side of the
metal--insulator transition, the electric field of a charged dot is
screened and the energy (19) is insignificant. On the insulator side there
is no screening of free carriers, and the number of charged dots is
determined by comparing the energy (19) with the temperature. When
$\varepsilon_e\ll T$, the number $\nu$ of charged dots is exponentially
small:
\begin{equation}\label{nu}
  \nu=(n/N)\exp(-\varepsilon_e/T).
\end{equation}
Since conduction in these conditions is determined by tunnelling between
charged and uncharged dots, $\nu$ acts as the number of carriers. This
standard line of reasoning used to describe granulated metals [28]
determines the activation nature of the conduction.

The importance of replacing $n$ with $n/N$ can be illustrated by the fact
that when an amorphous alloy is transformed by annealing into a
quasicrystal, its resistivity often increases severalfold [18].
Localization is also substantially enhanced by the absence of
translational symmetry and of universal short-range order in the mutual
arrangement of configurations. Irregularities in this mutual arrangement
prevents resonant tunnelling. Of course, the presence of translational
symmetry by itself cannot guarantee metallic conduction. Is is obvious
from Fig. 1, where a Mott transition is possible even at $W=0$. Near the
limiting values of concentrations, $n\gtrsim n_{\rm Mott}$, disorder is
essential. Indeed, when the CsAu alloy crystallizes, its resistivity
decreases by a factor of 10 (see Fig. 10).

Thus, the two classes of condensed media briefly discussed in this paper
provide an affirmative answer to the question posed at the end of the
Introduction: {\it It is possible to localize a system of valence
electrons in a medium consisting only of metal atoms.} Such localization
is realized through the formation of molecule-like configurations at least
in two cases: in two-component melts with an alkali metal as one of the
components, and in quasicrystals.

\end{document}